\documentclass[11pt,superscriptaddress,aps,prd,preprint]{revtex4}

\everymath{\displaystyle}
\usepackage{graphicx}

\usepackage{amsmath,amssymb,mathrsfs}

\newcommand{\bea}{\begin{eqnarray}}
\newcommand{\eea}{\end{eqnarray}}

\begin{document}

\title{On the Gravitational Casimir Effect at Finite Temperature for Gravitational Waves}

\author{S. C. Ulhoa} 
\email{sc.ulhoa@gmail.com}
\affiliation{Instituto de F\'{\i}sica, Universidade de Bras\'{\i}lia, 70910-900, Bras\'{\i}lia, DF, Brazil}
\affiliation{Canadian Quantum Research Center, 106-460 Doyle Ave, Kelowna, British Columbia V1Y 0C2, Canada}

\author{A. F. Santos} 
\email{alesandroferreira@fisica.ufmt.br}
\affiliation{Programa de P\'{o}s-Gradua\c{c}\~{a}o em F\'{\i}sica, Instituto de F\'{\i}sica,\\ 
Universidade Federal de Mato Grosso, Cuiab\'{a}, Brasil}

\author{R. G. G. de Amorim} 
\email{ronniamorim@gmail.com}
\affiliation{Faculdade de Ci\^{e}ncias e Tecnologias em Engenharia, Universidade de Bras\'{\i}lia, 70910-900, Bras\'{\i}lia, DF, Brazil}
\affiliation{Canadian Quantum Research Center, 106-460 Doyle Ave, Kelowna, British Columbia V1Y 0C2, Canada}

\author{G. H. S. Vasconcelos}\email{lendohqs@gmail.com}
\affiliation{Instituto de F\'{\i}sica, Universidade de Bras\'{\i}lia, 70910-900, Bras\'{\i}lia, DF, Brazil}


\begin{abstract}
The article deals with the gravitational Casimir effect in a
weak plane-wave background. The calculation combines the
gravitational energy--momentum tensor of the teleparallel
equivalent of general relativity with a Green-function
prescription based on Thermo Field Dynamics. At zero temperature,
the stress tensor is evaluated for three spatial compactifications:
one along the propagation axis and one along each transverse
axis. To linear order in the wave amplitudes, the difference
between the transverse mean pressures at the same coordinate
period selects the $+$ polarization. The longitudinal mean
pressure is insensitive to the wave at this order, and the
$\times$ polarization enters the off-diagonal stresses rather
than the mean pressure. The finite-temperature calculation
is restricted to longitudinal compactification. In this case,
neither polarization enters the energy density at linear order,
and the sum of the thermal, vacuum Casimir, and mixed terms
crosses from negative to positive at $Td \simeq 0.32893$,
where $T$ is the temperature and $2d$ is the spatial period
in natural units.
\end{abstract}

\maketitle

\date{\today}

\section{Introduction}
\label{sec.1}

The Casimir effect illustrates how boundary conditions and spatial
topology affect the vacuum energy and stresses of a quantum field.
Originally derived for the electromagnetic field between ideal
conducting plates \cite{Casimir1948}, it has since been investigated
for different fields, geometries, and temperatures
\cite{Bordag2009}. In gravitational settings, these investigations
also address the relation between vacuum fluctuations and spacetime
geometry, including the combined effects of spatial compactification,
curvature, and temperature \cite{Bezerra2024}. These developments
motivate the question of how compactification and temperature affect
the energy and stresses of the gravitational field itself.

The gravitational Casimir effect has been approached from different
perspectives. Quach investigated Casimir forces associated with
quantized gravitational waves
\cite{Quach2015,Quach2017Erratum}, while Santos and Khanna studied
gravitational Casimir energy and pressure at finite temperature
within gravitoelectromagnetism \cite{SantosKhanna2016}.
A related but distinct line of investigation concerns the response
of other quantum fields to gravitational waves. Sorge examined
gravitational memory in the Casimir effect \cite{Sorge2023},
and de Oliveira \emph{et al.} studied particle production in a
cavity with a mirror driven by a gravitational wave
\cite{Oliveira2026}. In the present work, we consider the
compactification-dependent expectation value of the gravitational
energy--momentum tensor itself.

For this purpose, we adopt the teleparallel equivalent of general
relativity (TEGR), in which gravitational dynamics are formulated
in terms of tetrads and torsion. Although dynamically equivalent
to Einstein's formulation, TEGR provides an observer-dependent
expression for gravitational energy and momentum that can be
evaluated in finite spatial regions \cite{Maluf2013}.
This formulation also permits the study of the local energy and
stresses of gravitational waves; in particular, its energy--momentum
tensor for linearized waves has been shown to agree with the
usual general-relativistic expression \cite{Formiga2018}.
Here, we use the TEGR tensor to calculate the energy density
and stresses of a weak gravitational wave under spatial
compactification and at finite temperature.

We evaluate these quantities using Thermo Field Dynamics (TFD),
a real-time formalism in which thermal averages are represented
as expectation values in a thermal vacuum in a doubled Hilbert
space \cite{TakahashiUmezawa1975}. Generalized Bogoliubov
transformations extend this construction to spatial
compactification, allowing thermal and spatial periodicities
to be incorporated within the same Green-function prescription
\cite{Silva2002,Khanna2014}. The combination of TEGR and TFD
has already been used to derive gravitational analogues of the
Stefan--Boltzmann law and the Casimir effect in nearly flat
spacetime \cite{Ulhoa2019}, as well as to investigate related
thermal and Casimir quantities in de Sitter spacetime
\cite{Santos2021}.

A plane gravitational wave introduces a directional structure
that allows these studies to be extended. Its propagation axis
is distinguished from the transverse plane, while its
polarizations distinguish the perturbations within that plane.
This raises the question of how the gravitational Casimir
stress depends on the orientation of the compactified direction
relative to the wave. To investigate this dependence, we
consider a weak plane wave and compare, at zero temperature,
compactification along the propagation axis and along each
of the two transverse axes, keeping the same coordinate
period. The comparison identifies how the two polarizations
enter the gravitational stresses and how the difference
between the transverse mean pressures selects the
$+$ contribution.

We then introduce temperature while retaining spatial
compactification along the propagation direction, and turn
from the stresses to the energy density of the wave.
In this configuration, the energy density has no explicit
polarization dependence at first order in the wave amplitudes,
but changes sign as the temperature increases at fixed
compactification period. The two calculations thus examine
complementary aspects of the gravitational energy--momentum
tensor: its response to the orientation of spatial
compactification and the thermal behavior of its energy
density.

The paper is organized as follows.
Section~\ref{thermo} introduces TFD and the generalized
compactification prescription.
Section~\ref{sec.2} reviews the TEGR formulation of gravitational
energy and momentum.
Section~\ref{sec.3} presents the zero-temperature comparison
of compactification directions and the finite-temperature
calculation for longitudinal compactification.
Section~\ref{sec.4} summarizes the results and discusses
possible extensions.

\section{Thermo Field Dynamics (TFD)}
\label{thermo}

Thermo Field Dynamics (TFD) is a thermal quantum field theory that is
used when it is desirable to keep an explicit time dependence in
addition to the temperature. It is based on two ingredients. The first
is the doubling of the Hilbert space $\mathcal S$ of the original field
system, which gives rise to
\begin{equation}
\mathcal S_T
=
\mathcal S\otimes\widetilde{\mathcal S},
\label{eq:tfd-hilbert-space}
\end{equation}
where $\widetilde{\mathcal S}$ is the tilde (dual) space; the doubling
is defined by the tilde conjugation rules. The second is the Bogoliubov
transformation, which introduces the thermal effects through a rotation
between tilde and non-tilde operators. Together, these ingredients allow
the statistical average of an arbitrary operator $A$ to be interpreted
as an expectation value in a thermal vacuum,
\begin{equation}
\langle A\rangle_\beta
=
\frac{\operatorname{Tr}(e^{-\beta H}A)}
     {\operatorname{Tr}(e^{-\beta H})}
=
\langle 0(\beta)|A|0(\beta)\rangle,
\label{eq:tfd-thermal-average}
\end{equation}
where $H$ is the Hamiltonian of the system and the thermal vacuum
$|0(\beta)\rangle$ describes its thermal equilibrium at the temperature
$T$, with $\beta=1/T$ in units in which the Boltzmann constant is equal
to one.

For a bosonic field with annihilation and creation operators $a(k)$ and
$a^\dagger(k)$, and tilde counterparts $\widetilde a(k)$ and
$\widetilde a^\dagger(k)$, the Bogoliubov transformation reads
\begin{equation}
\begin{pmatrix}
a(k;\alpha)\\
\widetilde a^\dagger(k;\alpha)
\end{pmatrix}
=
\begin{pmatrix}
u(k;\alpha)&-v(k;\alpha)\\
-v(k;\alpha)&u(k;\alpha)
\end{pmatrix}
\begin{pmatrix}
a(k)\\
\widetilde a^\dagger(k)
\end{pmatrix},
\label{eq:tfd-bogoliubov}
\end{equation}
with
\begin{equation}
u^2(k;\alpha)-v^2(k;\alpha)=1.
\label{eq:tfd-bosonic-condition}
\end{equation}
Here $k$ is the four-momentum and
$\alpha=(\alpha_0,\alpha_1,\alpha_2,\alpha_3)$ is the compactification
parameter. The temperature effect is described by the choice
$\alpha=(\beta,0,0,0)$, for which
\begin{equation}
v^2(k^0;\beta)
=
\frac{1}{e^{\beta k^0}-1}
=
\sum_{m=1}^{\infty}e^{-m\beta k^0},
\label{eq:tfd-thermal-coefficient}
\end{equation}
where $k^0>0$.

The parameter $\alpha$ is not restricted to the temperature: its
remaining components describe the compactification of the spatial
coordinates, which is the choice suited to the Casimir effect. For the
compactification of the $x^3$ direction, $\alpha=(0,0,0,i2d)$ and
\begin{equation}
v^2(k^3;d)
=
\sum_{l=1}^{\infty}e^{-i2ld k^3},
\label{eq:tfd-spatial-coefficient}
\end{equation}
where the length $d$ fixes the period $2d$ of the compactification.
The other spatial directions are treated in the same way. Temperature
and spatial compactification are considered simultaneously with the
choice
\begin{equation}
\alpha=(\beta,0,0,i2d).
\label{eq:tfd-compactification}
\end{equation}
In this case the generalized Bogoliubov transformation is
\begin{equation}
v^2(k^0,k^3;\beta,d)
=
v^2(k^0;\beta)
+
v^2(k^3;d)
+
2v^2(k^0;\beta)\,v^2(k^3;d),
\label{eq:tfd-combined-coefficient}
\end{equation}
that is,
\begin{equation}
v^2(k^0,k^3;\beta,d)
=
\sum_{m=1}^{\infty}e^{-m\beta k^0}
+
\sum_{l=1}^{\infty}e^{-i2ld k^3}
+
2\sum_{m,l=1}^{\infty}
e^{-m\beta k^0-i2ld k^3},
\label{eq:tfd-combined-series}
\end{equation}
where the first term takes into account the temperature effects, the
second stands for the spatial compactification only, and the last one
for the interaction between both.

The compactification parameter enters the physical quantities through
the Green function of the field. In the doubled notation the Green
function is a $2\times2$ matrix, $G_0^{(AB)}(x-x';\alpha)$ with
$A,B=1,2$, and the physical quantities are given by the non-tilde
variables, i.e., by the component $G_0^{(11)}(x-x';\alpha)$. In momentum
space, this component differs from the Green function $G_0$ of the field
in the absence of compactification by a term proportional to
$v^2(k;\alpha)$. It is then convenient to define
\begin{equation}
\overline G_0(x-x';\alpha)
=
G_0^{(11)}(x-x';\alpha)-G_0(x-x'),
\label{eq:tfd-green-difference}
\end{equation}
which carries the whole dependence on $\alpha$. Since each term of the
series in Eqs.~(\ref{eq:tfd-thermal-coefficient}),
(\ref{eq:tfd-spatial-coefficient}), and (\ref{eq:tfd-combined-series})
is an exponential in the components of the momentum, its effect on
$G_0(x-x')$ is a displacement of the argument. For
$\alpha=(\beta,0,0,0)$ one obtains
\begin{equation}
\overline G_0(x-x';\beta)
=
2\sum_{m=1}^{\infty}
G_0(x-x'-im\beta\,n_0),
\label{eq:tfd-thermal-images}
\end{equation}
and for $\alpha=(0,0,0,i2d)$,
\begin{equation}
\overline G_0(x-x';d)
=
2\sum_{l=1}^{\infty}
G_0(x-x'-2ld\,n_3),
\label{eq:tfd-spatial-images}
\end{equation}
where $n_0=(1,0,0,0)$ and $n_3=(0,0,0,1)$. For the choice of
Eq.~(\ref{eq:tfd-compactification}), the three terms of
Eq.~(\ref{eq:tfd-combined-series}) lead to
\begin{equation}
\begin{aligned}
\overline G_0(x-x';\beta,d)
={}&
2\sum_{m=1}^{\infty}
G_0(x-x'-im\beta\,n_0)
+
2\sum_{l=1}^{\infty}
G_0(x-x'-2ld\,n_3)
\\
&+
4\sum_{m,l=1}^{\infty}
G_0(x-x'-im\beta\,n_0-2ld\,n_3).
\end{aligned}
\label{eq:tfd-combined-images}
\end{equation}
In this form, the formalism requires only the characterization of the
field under analysis by means of the corresponding Green function $G_0$.

\section{Teleparallelism Equivalent to General Relativity (TEGR)} \label{sec.2}

The gravitational field to which the TFD formalism will be applied is
described by the teleparallel equivalent of general relativity (TEGR),
which we briefly review in this section, following Ref.~\cite{Maluf2013}.
The fundamental
dynamical variables of the theory are the tetrad fields
$e^a{}_\mu$, which encode the gravitational field and define
orthonormal frames that can be adapted to specific observers.
The tetrad formulation thus provides a unified description of the
gravitational field and the reference frames used to describe it.
This dual role is reflected in the two distinct types of indices
carried by the tetrad: Latin indices refer to the tangent-space
Lorentz frame, whereas Greek indices refer to spacetime coordinates.
These indices distinguish the action of Lorentz transformations
on the frame from that of coordinate transformations on spacetime.

In the standard Riemannian formulation of general relativity,
the geometrical description of the gravitational field is based
on the Christoffel symbols ${}^{0}\Gamma_{\mu\lambda\nu}$.
However, the same gravitational dynamics can be described within
an alternative geometrical framework, giving rise to the
teleparallel equivalent of general relativity. In the formulation
adopted here, the tetrad fields define the Weitzenb\"ock connection,
\begin{equation}
\Gamma_{\mu\lambda\nu}
=
e^a{}_\mu\,\partial_\lambda e_{a\nu},
\label{eq:weitzenbock-connection}
\end{equation}
which has vanishing curvature and, in general, nonvanishing torsion.
The torsion tensor associated with this connection is expressed
in terms of the tetrad fields as
\begin{equation}
T^a{}_{\lambda\nu}
=
\partial_\lambda e^a{}_\nu
-
\partial_\nu e^a{}_\lambda.
\label{4}
\end{equation}
The Weitzenb\"ock connection and the Christoffel symbols are
related by the identity
\begin{equation}
\Gamma_{\mu\lambda\nu}
=
{}^0\Gamma_{\mu\lambda\nu}
+
K_{\mu\lambda\nu},
\label{2}
\end{equation}
where $K_{\mu\lambda\nu}$ is the contortion tensor,
\begin{equation}
K_{\mu\lambda\nu}
=
\frac{1}{2}
\left(
T_{\lambda\mu\nu}
+
T_{\nu\lambda\mu}
+
T_{\mu\lambda\nu}
\right),
\label{3}
\end{equation}
and $T_{\mu\lambda\nu}=e_{a\mu}T^a{}_{\lambda\nu}$.

Equation~(\ref{2}) establishes the geometrical relation underlying
the equivalence between the teleparallel and Riemannian
formulations. In particular, the vanishing of the curvature
constructed from $\Gamma_{\mu\lambda\nu}$ allows the scalar
curvature density constructed from the Christoffel symbols
to be rewritten entirely in terms of the torsion tensor and
a total divergence:
\begin{equation}
eR(e)
\equiv
-e\left(
\frac{1}{4}T^{abc}T_{abc}
+
\frac{1}{2}T^{abc}T_{bac}
-
T^aT_a
\right)
+
2\partial_\mu(eT^\mu),
\label{eq5}
\end{equation}
where $e=\det(e^a{}_\mu)$,
$T^a=T^b{}_b{}^a$,
$T_\mu=T^\nu{}_{\nu\mu}$, and
$T^\mu=g^{\mu\nu}T_\nu$.
The scalar $R(e)$ is the Riemannian scalar curvature associated
with the metric determined by the tetrad fields.

The identity~(\ref{eq5}) shows that the Hilbert--Einstein
Lagrangian density and a particular quadratic combination
of the torsion tensor differ by a total divergence.
This leads to the TEGR Lagrangian density
\begin{equation}
\begin{aligned}
\mathfrak L(e_{a\mu})
={}&
-\kappa e
\left(
\frac{1}{4}T^{abc}T_{abc}
+
\frac{1}{2}T^{abc}T_{bac}
-
T^aT_a
\right)
-
\mathfrak L_M
\\
\equiv{}&
-\kappa e\,\Sigma^{abc}T_{abc}
-
\mathfrak L_M,
\end{aligned}
\label{6}
\end{equation}
where $\mathfrak L_M$ denotes the matter Lagrangian density,
$\kappa=1/(16\pi)$ in units $G=c=1$, and
\begin{equation}
\Sigma^{abc}
=
\frac{1}{4}
\left(
T^{abc}+T^{bac}-T^{cab}
\right)
+
\frac{1}{2}
\left(
\eta^{ac}T^b-\eta^{ab}T^c
\right).
\label{7}
\end{equation}
It should be noted that the total divergence in Eq.~(\ref{eq5})
has been omitted in defining the Lagrangian density~(\ref{6}),
since it does not affect the field equations under the appropriate
boundary conditions. Its omission, however, removes the invariance
of the Lagrangian density under arbitrary local Lorentz
transformations. In this formulation, the gravitational
energy-momentum and angular momentum are covariant under global,
but not local, Lorentz transformations, reflecting their dependence
on the reference frame. This does not imply a loss of local Lorentz
covariance at the level of the dynamics: the field equations remain
covariant under local Lorentz transformations and are equivalent
to Einstein's equations~\cite{Maluf2013}.

The variation of the Lagrangian density~(\ref{6}) with respect
to the tetrad fields yields the field equations, which can be written
in the form
\begin{equation}
\partial_\nu\left(e\Sigma^{a\lambda\nu}\right)
=
\frac{1}{4\kappa}
e\,e^a{}_\mu
\left(
t^{\lambda\mu}+T^{\lambda\mu}
\right),
\label{10}
\end{equation}
where $T^{\lambda\mu}$ is the energy-momentum tensor of matter fields
and
\begin{equation}
t^{\lambda\mu}
=
\kappa
\left[
4\Sigma^{bc\lambda}T_{bc}{}^\mu
-
g^{\lambda\mu}\Sigma^{abc}T_{abc}
\right].
\label{11}
\end{equation}
Since $t^{\lambda\mu}$ and $T^{\lambda\mu}$ appear on the same footing
in Eq.~(\ref{10}), $t^{\lambda\mu}$ is interpreted as the gravitational
energy-momentum tensor. The antisymmetry of $\Sigma^{a\lambda\nu}$ in its last
two indices implies
\begin{equation}
\partial_\lambda\partial_\nu
\left(e\Sigma^{a\lambda\nu}\right)
\equiv0.
\label{12}
\end{equation}
Consequently, the field equations yield the local
conservation equation
\begin{equation*}
\partial_\lambda
\left[
e\,e^a{}_\mu
\left(
t^{\lambda\mu}+T^{\lambda\mu}
\right)
\right]
=
0.
\end{equation*}
This equation motivates the definition of the total
energy-momentum contained within a spatial volume $V$,
\begin{equation}
P^a
=
\int_V d^3x\,
e\,e^a{}_\mu
\left(
t^{0\mu}+T^{0\mu}
\right).
\label{14}
\end{equation}
Using the $\lambda=0$ component of Eq.~(\ref{10}),
the same quantity can be expressed as
\begin{equation}
P^a
=
4\kappa
\int_V d^3x\,
\partial_\nu
\left(e\Sigma^{a0\nu}\right).
\label{14.1}
\end{equation}
Since $\Sigma^{a00}=0$, only spatial derivatives contribute
to Eq.~(\ref{14.1}), allowing $P^a$ to be written as
a surface integral over the boundary of $V$.
In vacuum, this expression gives the gravitational
energy-momentum contained within that volume.

An important feature of this definition is that $P^a$ transforms
as a four-vector under global Lorentz transformations.
Consequently, its time component, which gives the gravitational
energy in vacuum, depends on the observer's reference frame,
as expected for the energy of a physical system.
This frame dependence is distinct from coordinate dependence:
the definition is invariant under transformations of the spatial
coordinates and under time reparametrizations.
Moreover, it can be evaluated within a finite spatial volume,
rather than only at spatial infinity.
Taken together, these properties support a physically consistent
definition of gravitational energy, combining coordinate
invariance with the observer dependence required by its
character as the time component of an energy-momentum
four-vector.

\section{Gravitational Casimir effect at zero and finite temperature}
\label{sec.3}

We now apply the TFD formalism of Sec.~\ref{thermo} to the gravitational
energy-momentum tensor $t^{\lambda\mu}$ of Eq.~(\ref{11}) in the
presence of a weak gravitational wave. We first consider spatial
compactification along each of the three coordinate directions at zero
temperature. We then introduce temperature for compactification along
the propagation direction of the wave.

In the coordinates $x^\mu=(t,x,y,z)$, the metric is written as
\begin{equation}
g_{\mu\nu}=\eta_{\mu\nu}+h_{\mu\nu},
\label{eq:weak-field-metric}
\end{equation}
where $\eta_{\mu\nu}=\operatorname{diag}(-1,1,1,1)$ and $h_{\mu\nu}$
describes a plane gravitational wave propagating along the $z$
direction. In the transverse-traceless gauge, the perturbation takes
the form
\begin{equation}
h_{\mu\nu}=
\begin{pmatrix}
0&0&0&0\\
0&h_+(u)&h_\times(u)&0\\
0&h_\times(u)&-h_+(u)&0\\
0&0&0&0
\end{pmatrix},
\label{eq:wave-polarizations}
\end{equation}
where $u=t-z$ and the functions $h_+$ and $h_\times$ represent the two
polarization states. We assume $|h_+|,|h_\times|\ll1$ and, throughout
this section, retain terms through first order in these amplitudes, so
that
\begin{equation}
g^{\mu\nu}
=
\eta^{\mu\nu}-h^{\mu\nu}+O(h^2).
\label{eq:inverse-wave-metric}
\end{equation}

In the weak-field approximation of the TEGR, the mean value of the
gravitational energy-momentum tensor is obtained from the Green function
$G_0$, with the tensor written at two different points, $x$ and $x'$,
and the limit $x'\to x$ taken afterwards. For the metric of
Eq.~(\ref{eq:weak-field-metric}), the Green function used in the
calculation is
\begin{equation}
G_0(x-x')
=
-\frac{i64\pi}
       {g_{\mu\nu}(x)q^\mu q^\nu},
\label{eq:gravitational-green-function}
\end{equation}
where $q^\mu=x^\mu-x'^\mu$. Since the metric coefficients depend on the
wave phase $u$, the Green function depends on $x$ also through
$g_{\mu\nu}(x)$; this dependence is not displayed in the argument of
$G_0$, but it is kept in the calculation. In TFD, $G_0$ is replaced by
the $\alpha$-dependent Green function of Sec.~\ref{thermo}, and the
Casimir prescription,
\begin{equation}
\mathcal T^{\mu\nu}(x;\alpha)
=
\langle t^{\mu\nu}(x;\alpha)\rangle
-
\langle t^{\mu\nu}(x)\rangle,
\label{eq:casimir-prescription}
\end{equation}
in which the mean value in the absence of compactification is
subtracted, leads to
\begin{equation}
\mathcal T^{\mu\nu}(x;\alpha)
=
4i\kappa\lim_{x'\to x}
\left[
-5g^{\mu\nu}g^{\rho\sigma}
+
2g^{\mu\rho}g^{\nu\sigma}
\right]
\partial'_\rho\partial_\sigma
\overline G_0(x-x';\alpha),
\label{eq:gravitational-energy-momentum}
\end{equation}
where $\overline G_0$ is defined in
Eq.~(\ref{eq:tfd-green-difference}), the metric is evaluated at
$x$, and $\partial_\sigma$ and $\partial'_\rho$ denote the derivatives
with respect to $x^\sigma$ and $x'^\rho$, respectively. The derivatives
are evaluated before taking the coincidence limit, including the
coordinate dependence of the metric in the Green function.

\subsection{Spatial compactification at zero temperature}

At zero temperature, the compactification of one spatial direction with
period $2d$ corresponds to $\alpha=(0,i2d,0,0)$, $\alpha=(0,0,i2d,0)$,
or $\alpha=(0,0,0,i2d)$ for the $x$, $y$, or $z$ direction,
respectively. The corresponding Green function is given by
Eq.~(\ref{eq:tfd-spatial-images}), with $n_3$ replaced by
$n_1=(0,1,0,0)$ in the first case and by $n_2=(0,0,1,0)$ in the second,
and with $G_0$ given by Eq.~(\ref{eq:gravitational-green-function}).

We evaluate Eq.~(\ref{eq:gravitational-energy-momentum}) separately for
the three cases. To simplify the notation, the compactification
direction is specified in the text rather than by an additional label
on $\mathcal T^{\mu\nu}$, and the dependence on $x$ through the wave
amplitudes is left implicit, so that the components are denoted by
$\mathcal T^{\mu\nu}(d)$. In each case we list the nonvanishing
components, with $\dot h_+\equiv\partial_t h_+$ and
$\dot h_\times\equiv\partial_t h_\times$.

For compactification along the $x$ direction, we obtain
\begin{equation}
\begin{aligned}
\mathcal T^{00}(d)
&=
\frac{4\pi^4}{45d^4}(2h_+-1),
\\
\mathcal T^{10}(d)
=
\mathcal T^{13}(d)
&=
-\frac{16\zeta(3)}{d^3}\dot h_+,
\\
\mathcal T^{11}(d)
&=
\frac{4\pi^4}{15d^4}(3h_+-1),
\\
\mathcal T^{12}(d)
=
\mathcal T^{21}(d)
&=
-\frac{4\pi^4}{45d^4}h_\times,
\\
\mathcal T^{20}(d)
=
\mathcal T^{23}(d)
&=
\frac{16\zeta(3)}{d^3}\dot h_\times,
\\
\mathcal T^{22}(d)
&=
\frac{4\pi^4}{45d^4}(1-h_+),
\\
\mathcal T^{33}(d)
&=
\frac{4\pi^4}{45d^4}(1-2h_+).
\end{aligned}
\label{eq:casimir-x}
\end{equation}
For compactification along the $y$ direction, the corresponding
components are
\begin{equation}
\begin{aligned}
\mathcal T^{00}(d)
&=
-\frac{4\pi^4}{45d^4}(1+2h_+),
\\
\mathcal T^{10}(d)
=
\mathcal T^{13}(d)
&=
\frac{16\zeta(3)}{d^3}\dot h_\times,
\\
\mathcal T^{11}(d)
&=
\frac{4\pi^4}{45d^4}(1+h_+),
\\
\mathcal T^{12}(d)
=
\mathcal T^{21}(d)
&=
-\frac{4\pi^4}{45d^4}h_\times,
\\
\mathcal T^{20}(d)
=
\mathcal T^{23}(d)
&=
\frac{16\zeta(3)}{d^3}\dot h_+,
\\
\mathcal T^{22}(d)
&=
-\frac{4\pi^4}{15d^4}(1+3h_+),
\\
\mathcal T^{33}(d)
&=
\frac{4\pi^4}{45d^4}(1+2h_+).
\end{aligned}
\label{eq:casimir-y}
\end{equation}
In both transverse compactifications, the energy density depends on
$h_+$, whereas $h_\times$ contributes to off-diagonal components. Terms
involving the derivatives of the polarization amplitudes also appear in
off-diagonal components; they originate from the coordinate dependence
of the metric in the Green function.

For compactification along the $z$ direction, parallel to the
propagation of the wave, we find
\begin{equation}
\begin{aligned}
\mathcal T^{00}(d)
&=
-\frac{4\pi^4}{45d^4},
\\
\mathcal T^{11}(d)
&=
\frac{4\pi^4}{45d^4}(1-h_+),
\\
\mathcal T^{12}(d)
=
\mathcal T^{21}(d)
&=
-\frac{4\pi^4}{45d^4}h_\times,
\\
\mathcal T^{22}(d)
&=
\frac{4\pi^4}{45d^4}(1+h_+),
\\
\mathcal T^{33}(d)
&=
-\frac{4\pi^4}{15d^4}.
\end{aligned}
\label{eq:casimir-z}
\end{equation}
Unlike the transverse cases, the energy density and the longitudinal
stress receive no first-order correction from the wave amplitudes, and
no terms involving their derivatives appear. The polarization
dependence remains in the transverse spatial components.

For each compactification, we define the mean gravitational pressure by
\begin{equation}
\bar p(d)
=
\frac{1}{3}g_{jk}\mathcal T^{jk}(d),
\label{eq:mean-casimir-pressure}
\end{equation}
where the spatial indices $j,k=1,2,3$ are summed. With the metric of
Eq.~(\ref{eq:weak-field-metric}), this contraction reads
\begin{equation}
3\bar p(d)
=
(1+h_+)\mathcal T^{11}(d)
+
(1-h_+)\mathcal T^{22}(d)
+
2h_\times\mathcal T^{12}(d)
+
\mathcal T^{33}(d).
\label{eq:mean-pressure-contraction}
\end{equation}
The $\times$ polarization enters only through the term
$2h_\times\mathcal T^{12}(d)$, which is of second order in the wave
amplitudes, since $\mathcal T^{12}(d)$ is proportional to $h_\times$
in the three cases. Its contribution to the mean pressure is therefore
part of the second-order analysis, whereas the expressions below
describe the first-order polarization dependence. Substituting the
components of Eqs.~(\ref{eq:casimir-x}), (\ref{eq:casimir-y}),
and (\ref{eq:casimir-z}) into
Eq.~(\ref{eq:mean-pressure-contraction}) and retaining terms through
first order in the wave amplitudes, we obtain
\begin{equation}
\begin{aligned}
\bar p_x(d)
&=
-\frac{4\pi^4}{135d^4}(1-2h_+),
\\
\bar p_y(d)
&=
-\frac{4\pi^4}{135d^4}(1+2h_+),
\\
\bar p_z(d)
&=
-\frac{4\pi^4}{135d^4},
\end{aligned}
\label{eq:mean-casimir-pressure-directions}
\end{equation}
where the subscript indicates the compactification direction.

The mean pressure therefore depends on the orientation of the
compactification relative to the propagation direction of the
gravitational wave. The two transverse configurations acquire
corrections of opposite signs, proportional to $h_+(u)$, while the
longitudinal configuration retains its unperturbed value at this order.
At fixed $d$, the change associated with the passage of the wave is
\begin{equation}
\delta\bar p(d)
\equiv
\bar p(d)
-
\left.\bar p(d)\right|_{h_+=h_\times=0}.
\label{eq:wave-induced-mean-pressure}
\end{equation}
For the three compactification directions, this gives
\begin{equation}
\begin{aligned}
\delta\bar p_x(d)
&=
\frac{8\pi^4}{135d^4}h_+(u),
\\
\delta\bar p_y(d)
&=
-\frac{8\pi^4}{135d^4}h_+(u),
\\
\delta\bar p_z(d)
&=0,
\end{aligned}
\label{eq:wave-induced-pressure-directions}
\end{equation}
to first order. In particular, the comparison of the two transverse
configurations yields
\begin{equation}
\delta\bar p_x(d)-\delta\bar p_y(d)
=
\frac{16\pi^4}{135d^4}h_+(u).
\label{eq:transverse-pressure-difference}
\end{equation}
This differential response characterizes the influence of the $+$
polarization on the mean pressure.

An experimental platform relevant to this result was developed by
Norte et al.~\cite{Norte2018}, who implemented an on-chip
optomechanical device for measuring changes in the Casimir force
between two closely spaced superconducting surfaces. In this setup, a
force-induced displacement of a movable element is read out through a
shift in the optical cavity resonance. This approach provides an
experimental basis for exploring small additional forces in Casimir
configurations.

The orientation dependence obtained here motivates a possible
application of this type of platform to the search for a gravitational
Casimir contribution associated with the passage of a gravitational
wave. A prospective arrangement would compare two otherwise identical
configurations whose plate normals, which correspond to the
compactification directions, are aligned with the transverse $x$ and
$y$ directions. According to
Eq.~(\ref{eq:wave-induced-pressure-directions}), the corresponding mean
pressures acquire opposite corrections proportional to $h_+(u)$,
yielding the differential response of
Eq.~(\ref{eq:transverse-pressure-difference}). A third configuration,
with its plate normal aligned with the propagation direction, would
provide the longitudinal comparison. These results motivate the
development of an optomechanical measurement scheme relating the
orientation-dependent gravitational stresses to the mechanical response
of the sensing elements, with the aim of probing the wave-induced
Casimir effect.

\subsection{Spatial compactification at finite temperature}

We now introduce temperature, keeping the spatial compactification
along the propagation direction of the wave. This is the choice of
Eq.~(\ref{eq:tfd-compactification}), for which the Green function is
given by Eq.~(\ref{eq:tfd-combined-images}), with $G_0$ of
Eq.~(\ref{eq:gravitational-green-function}). It contains a purely thermal
term, a zero-temperature spatial term, and a mixed term.

In these three terms, the argument of $G_0$ is displaced along $n_0$,
along $n_3$, and along both, respectively. Since $n_0$ and $n_3$ have
only temporal and longitudinal components, while the metric
perturbation has only transverse components, the displacements satisfy
\begin{equation}
g_{\mu\nu}
\left(im\beta\,n_0^\mu+2ld\,n_3^\mu\right)
\left(im\beta\,n_0^\nu+2ld\,n_3^\nu\right)
=
m^2\beta^2+4l^2d^2
\label{eq:combined-image-separation}
\end{equation}
and
\begin{equation}
(\partial_\sigma g_{\mu\nu})\,n_0^\nu
=
(\partial_\sigma g_{\mu\nu})\,n_3^\nu
=0.
\label{eq:longitudinal-metric-derivatives}
\end{equation}
These relations simplify the evaluation of the energy density, which
has no explicit dependence on the wave amplitudes at the order
considered.

Following the three terms of Eq.~(\ref{eq:tfd-combined-images}), we
write
\begin{equation}
\mathcal T^{00}(\beta,d)
=
\mathcal T^{00}(\beta)
+
\mathcal T^{00}(d)
+
\Delta\mathcal T^{00}(\beta,d),
\label{eq:energy-decomposition}
\end{equation}
where the first term is purely thermal, the second is the
zero-temperature Casimir contribution, and
$\Delta\mathcal T^{00}(\beta,d)$ denotes the mixed contribution.
Their explicit expressions are
\begin{equation}
\begin{aligned}
\mathcal T^{00}(\beta)
&=
\frac{384}{\beta^4}
\sum_{m=1}^{\infty}\frac{1}{m^4}
=
\frac{64\pi^4}{15\beta^4},
\\
\mathcal T^{00}(d)
&=
-\frac{8}{d^4}
\sum_{l=1}^{\infty}\frac{1}{l^4}
=
-\frac{4\pi^4}{45d^4},
\\
\Delta\mathcal T^{00}(\beta,d)
&=
256\sum_{m,l=1}^{\infty}
\frac{3m^2\beta^2-4l^2d^2}
     {(m^2\beta^2+4l^2d^2)^3},
\end{aligned}
\label{eq:energy-three-sectors}
\end{equation}
where $\zeta(4)=\pi^4/90$ has been used in the single sums. The thermal
term is positive and depends only on temperature, whereas the Casimir
term is negative, depends only on the compactification length, and
coincides with the energy density of Eq.~(\ref{eq:casimir-z}). The
double sum incorporates the simultaneous dependence on both scales.

To express this dependence in dimensionless form, we define
\begin{equation}
\begin{aligned}
\gamma&=Td=\frac{d}{\beta},
\\
\rho(\gamma)&=d^4\mathcal T^{00}(\beta,d).
\end{aligned}
\label{eq:dimensionless-variables}
\end{equation}
Equations~(\ref{eq:energy-decomposition}) and
(\ref{eq:energy-three-sectors}) then give
\begin{equation}
\rho(\gamma)
=
\frac{64\pi^4}{15}\gamma^4
-
\frac{4\pi^4}{45}
+
256\sum_{m,l=1}^{\infty}
\frac{3m^2\gamma^4-4l^2\gamma^6}
     {(m^2+4l^2\gamma^2)^3}.
\label{eq:rho-gamma}
\end{equation}
At fixed $d$, the zero-temperature limit recovers the longitudinal
Casimir energy density of Eq.~(\ref{eq:casimir-z}),
\begin{equation}
\lim_{\gamma\to0^+}\rho(\gamma)
=
-\frac{4\pi^4}{45}.
\label{eq:zero-temperature-limit}
\end{equation}

\begin{figure}[htbp]
\centering
\includegraphics[width=0.85\linewidth]{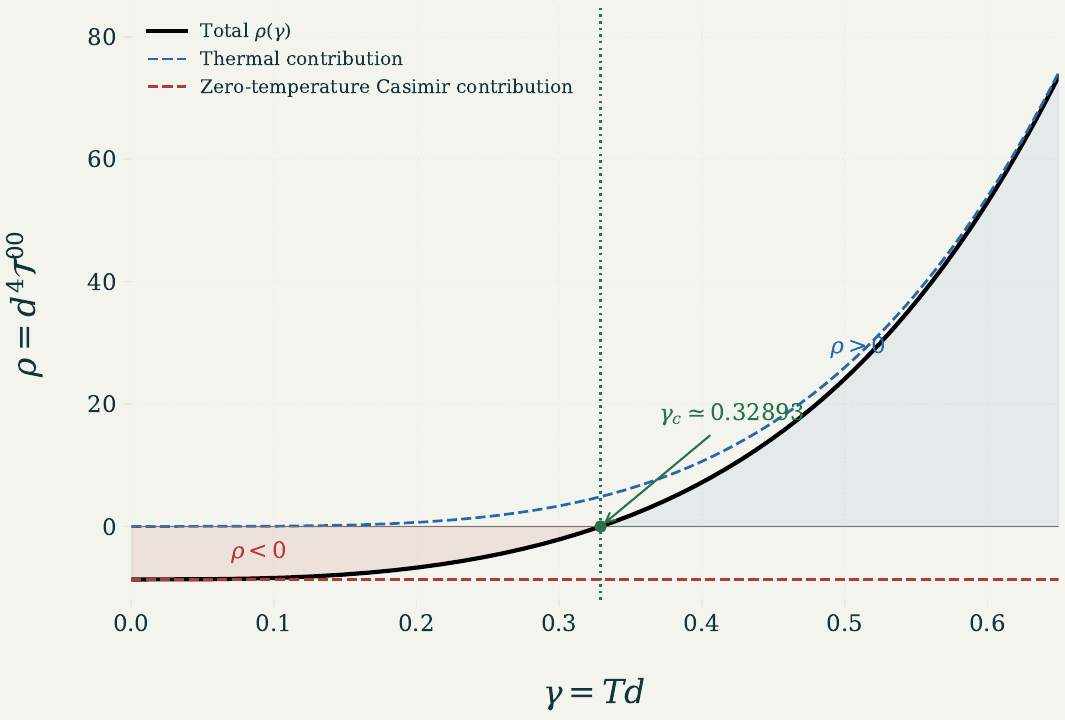}
\caption{
Dimensionless gravitational energy density
$\rho=d^4\mathcal T^{00}$ as a function of $\gamma=Td$
for spatial compactification along the propagation direction
of the gravitational wave.
The solid curve includes the thermal, zero-temperature Casimir,
and mixed contributions.
The dashed curves show the thermal and zero-temperature
Casimir contributions separately.
Both axes are linear.
The double sum is evaluated with $1\leq m,l\leq400$.
The marked zero occurs at $\gamma_c\simeq0.32893$.
}
\label{fig:rho-gamma}
\end{figure}

Figure~\ref{fig:rho-gamma} shows the dimensionless energy density
together with the purely thermal and zero-temperature Casimir
contributions. For the numerical truncation used in the figure, the
total energy density changes sign at $\gamma_c\simeq0.32893$. This zero
refers to the complete energy density, including the mixed
contribution, rather than to a cancellation of the two separate
single-sum contributions alone. The finite-temperature result thus
exhibits a temperature-driven change in sign of the gravitational
energy density: for a fixed compactification length $d$, the critical
temperature $T_c=\gamma_c/d$ separates the negative-energy regime from
the positive-energy regime.

A relevant consequence of the results obtained from the conventions and weak-field approximation adopted here is the differential pressure, given by 
\begin{equation}
\delta\bar p_x(d)-\delta\bar p_y(d)
 = \frac{16\pi^4}{135d^4}\,h_+(t-z)
 + \mathcal{O}(h^2).
\label{eq:conclusion_differential_pressure}
\end{equation}

In this relation it isolates an orientation-dependent contribution associated with the $+$ polarization. The $\times$ polarization, in turn, enters the off-diagonal stresses at first order, while its contribution to the scalar mean pressure arises only beyond that order. Accordingly, the full stress tensor contains polarization information that would not be apparent from the mean pressure alone.

\section{Conclusion} \label{sec.4}

We have investigated the gravitational Casimir effect in the
presence of a weak plane gravitational wave within the teleparallel
equivalent of general relativity (TEGR). Using Thermo Field Dynamics
(TFD) to implement spatial compactification and finite temperature,
we evaluated the compactification-dependent expectation value of
the gravitational energy--momentum tensor. The plane-wave geometry
allowed us to distinguish the effects of the two transverse--traceless
polarizations and to compare compactification transverse and parallel
to the direction of propagation.

At zero temperature and fixed coordinate period, the $+$ polarization
produces equal and opposite first-order corrections to the mean
gravitational pressure for compactification along $x$ and $y$.
The difference between the two transverse mean pressures therefore
isolates the term proportional to $h_+$, as given by
Eq.~\eqref{eq:conclusion_differential_pressure}. Compactification
along the propagation direction, $z$, leaves the mean pressure
unchanged at this order. The $\times$ polarization contributes to
off-diagonal stresses but does not enter the mean pressure at first
order. The mean pressure thus captures the transverse contrast
associated with the $+$ polarization, whereas the off-diagonal
components retain information about the $\times$ polarization.

At finite temperature, we restricted the calculation to longitudinal
compactification. To first order in the wave amplitudes, the energy
density has no explicit polarization dependence. It nevertheless
changes sign as the temperature increases at fixed compactification
period. This crossing is a property of the total energy density,
which includes the purely thermal, zero-temperature Casimir, and
mixed contributions. It does not, by itself, imply a reversal of
the Casimir force.

The contrast between the two transverse configurations motivates
a differential optomechanical comparison. One possible arrangement
would use two nominally identical Casimir cavities with orthogonal
surface normals in the transverse plane. A third cavity with its
normal along the propagation direction could provide a longitudinal
comparison. The superconducting Casimir-force platform demonstrated
by Norte \emph{et al.}~\cite{Norte2018} provides a precedent for
optomechanical force readout. It does not, however, establish a
correspondence between the pressure calculated here and the force on
a cavity element. In particular, a differential force prediction
would require a model of the normal stress at material boundaries
and of how the wave changes the proper separation of the surfaces.

A quantitative assessment would therefore require a treatment of
realistic boundary conditions and material properties, together with
a mechanical response model that distinguishes the proposed Casimir
signal from direct gravitational-wave strain and estimates the
relevant noise. Within the theoretical framework, natural extensions
are to treat transverse compactification at finite temperature, to
compute the stress tensor at second order in the wave amplitudes,
and to examine the dependence of the results on the observer frame
in TEGR. These extensions would clarify which features of the
orientation-dependent stress can be related to a measurable force.

\begin{acknowledgments}
A.~F.~Santos is partially supported by the Brazilian National Council
for Scientific and Technological Development (CNPq), Grant
No.~312406/2023-1. 
\end{acknowledgments}

\section*{Data Availability}
No datasets were generated or analysed during the current study.

\end{document}